\documentclass{evn2004}
\usepackage{txfonts}
\begin{document}
   \title{An Approach Detecting the Event Horizon of SgrA*}

   \author{M. Miyoshi\thanks{E-mail: makoto.miyoshi@nao.ac.jp} 
          }

   \institute{National Astronomical Observatory Japan, Mitaka, Tokyo Japan, 181-8588
             }

   \abstract{
 Imaging the vicinity of black hole is one of the ultimate goals of VLBI astronomy. The closest massive black hole, SgrA*, located at Galactic center is the leading candidate for such observations. Combined with recent VLBI recording technique and sub-mm radio engineering, we now have the sufficient sensitivity for the observations. We here show performance simulations of sub-mm VLBI arrays for imaging SgrA*. An excellent image is obtained from a sub-mm VLBI array in the Southern hemisphere like the configuration of VLBA. We also note that even with a small array, we can estimate the shadow size and then the mass of black hole from visibility analysis. Now, if only constructing a sub-mm VLBI array in Southern hemisphere, we can unveil the black hole environments of SgrA*.}

   \maketitle
%

\section{Introduction}

 Imaging black hole systems is one of the final goals in VLBI astronomy. SgrA*, the massive black hole at Galactic center is the leading candidate for the research. Because SgrA* shows the largest apparent angular size among black hole candidates. The apparent Schwarzschild radius is estimated to $6~\mu$arcsecs. from the mass ($2.6\times10^6 M_{\odot}$, Ghez et al. 2000) and the distance (8kpc). The corresponding shadow of black hole is about $30~\mu$arcsecs. in diameter (Falcke et al. 2000).  Recent observations indicate the mass of SgrA* is $3.7-4.1\times10^6 M_{\odot}$ (Sch\"{o}del et al. 2002, Ghez et al. 2003). If accepted the new values, the size of the black hole shadow is more than $45~\mu$arcsecs. in diameter.
 New findings of rapid flarings of SgrA* from a few hours to 30 min at radio, infrared, and x-ray emissions (Miyazaki et al. 2003, Zhao et al. 2004, Genzel et al. 2003., Baganoff et al. 2001, Goldwurm et al. 2003, Porquet et al. 2003) mean that the structure of the black hole system of SgrA* will also change rapidly.  SgrA* has become an important source for investigating black hole environments.
 Previous cm - to mm- VLBI observations show that the scattering effects by surrounding plasma blurred 
the intrinsic images of SgrA* (Doeleman et al. 2001) and that sub-mm VLBI will unveil the true image. 

 At the next section we show imaging performances of three sub-mm VLBI array configurations for SgrA*.

   \begin{figure}
   \centering
   \vspace{260pt}
  \includegraphics{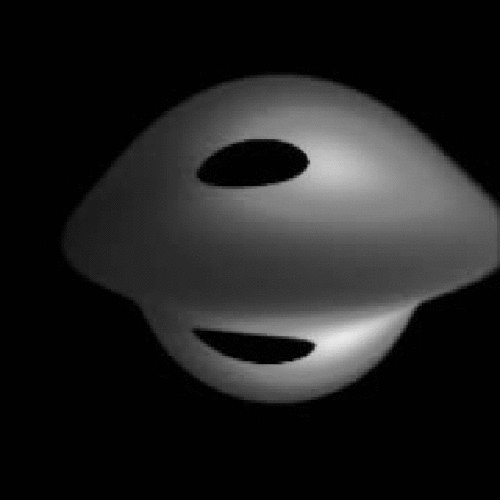}
      \caption{A image of SgrA* at 230~GHz in case of
 a Kerr hole with ADAF disk model calculated from the observed spectra of SgrA*.
 The viewing angle is assumed  to be $80 ^{\circ}$, almost edge-on.
The span of the figure is 12 Rs. This is from R. Takahashi in preparation. 
         }
   \end{figure}

\section{Imaging Performances of Three Array Configurations}
\subsection{Array configurations}
 We show three cases of array configurations. Array A is the same configuration as that of VLBA. Needless to say, the actual VLBA is only for below 86GHz observations. Array B consists of following five sub-mm telescopes (FST), namely the SMA in Hawaii, the CARMA in California, the ALMA, the SEST in Chile and a virtual station at Huancayo in Peru. The location of Huancayo is in lat.$12.0375 ^{\circ}$S. in long. $75.2942 ^{\circ}$W., 3375-m in altitude. Array C is the inverse VLBA. The antenna positions are inversed from the north to the south in latitude. We adopt the system temperatures around 80--150 K and the antenna efficiencies to be 0.6--0.7.
Figure 2 shows the uv-coverages of the three arrays for SgrA* at 230~GHz. The uv-coverage of the array A (VLBA) is notoriously worse in north-south direction (Fig. 2a). That of the array B is quite large but very sparse (Fig. 2b). The inverse VLBA shows dense uv-coverage (Fig. 2c). The corresponding synthesized beams (or dirty beams) are shown in Figure 3. 

\subsection{Image model for SgrA* at 230~GHz}
 The model image is a Gaussian shape with central shadow (Fig. 4a).
The  HPBW of the Gaussian is 0.1~mas in major axis and 0.08~mas in minor axis, the position angle is $80^{\circ}$. The central brightness is removed as the shadow (elliptical shape with $30~\mu$arcsecs. $\times 24~\mu$arcsecs.). The adopted outer size follows the first sub-mm VLBI detections of SgrA* (Krichbaum et al. 1998). We here assume the scattering effect is negligible at 230~GHz.

\subsection{Resultant Images from Clean}
 We performed Clean imaging with the task IMAGR in AIPS (NRAO). The restoring beams are unified to the circular Gaussian with the HPBW of $20~\mu$arcsecs. Figure 4 shows the resultant images.
 The images from the arrays A and B are not so good while the image from the array C has a dark area  at the center. The array C (inverse VLBA) shows the best image. Namely 8000~km in extent, 10 of stations, located at the Southern hemisphere are required for good imaging of SgrA*.

 The FST, a network of realistic sub-mm telescopes -- SMA, CARMA, SEST, ALMA and Huancayo-- does not show sufficient performance for the imaging. At following section we show the visibility analysis enable us to detect black hole shadow.

   \begin{figure*}
   \centering
   \vspace{200pt}
   \includegraphics{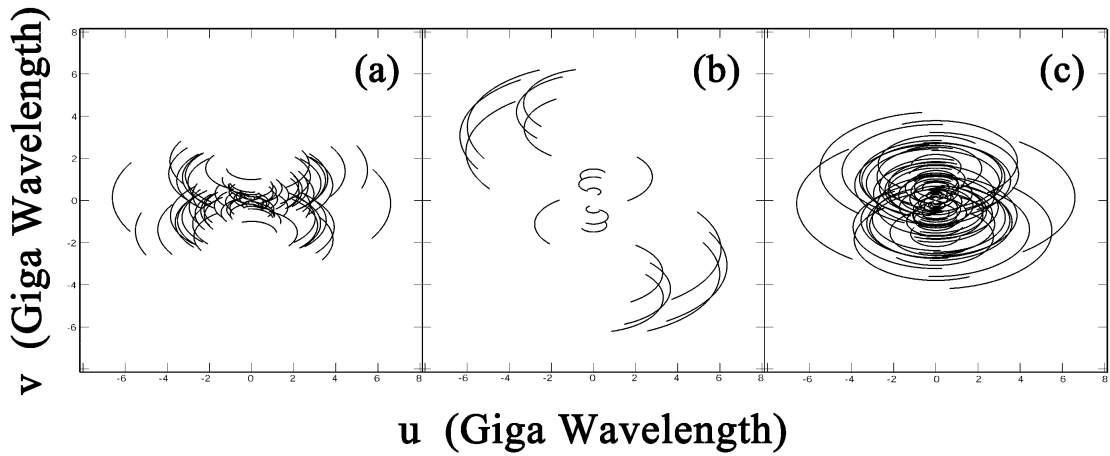}
   \caption{The uv coverages of the arrays for SgrA*. (a) Array A (VLBA), (b) Array B (FST) 
(c) Array C (inverse VLBA). The scale is 16 Giga wavelength in each side (230~GHz).
           }
    \end{figure*}
   \begin{figure*}
   \centering
   \vspace{200pt}
   \includegraphics{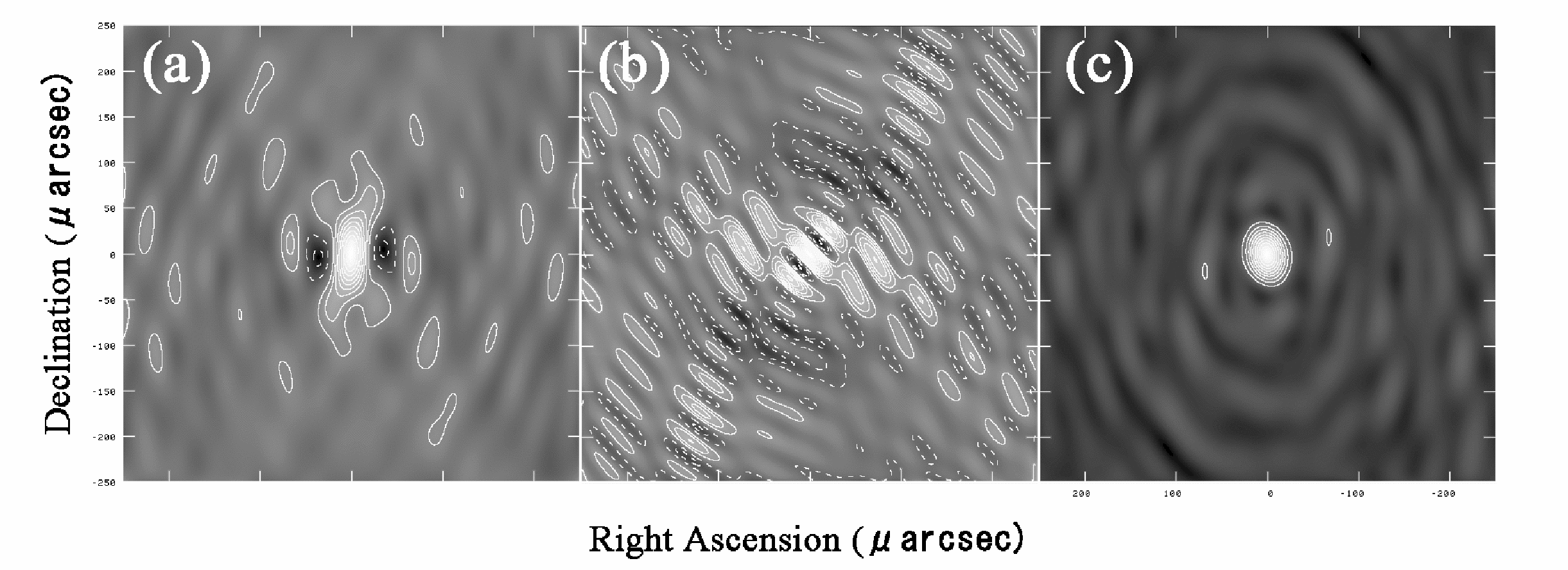}
   \caption{The corresponding synthesized beams. (a) Array A (VLBA),
 (b) Array B (FST) (c) Array C (inverse VLBA). The scale is 0.5mas in each side.
           }
    \end{figure*}

   \begin{figure*}
   \centering
   \vspace{170pt}
   \includegraphics{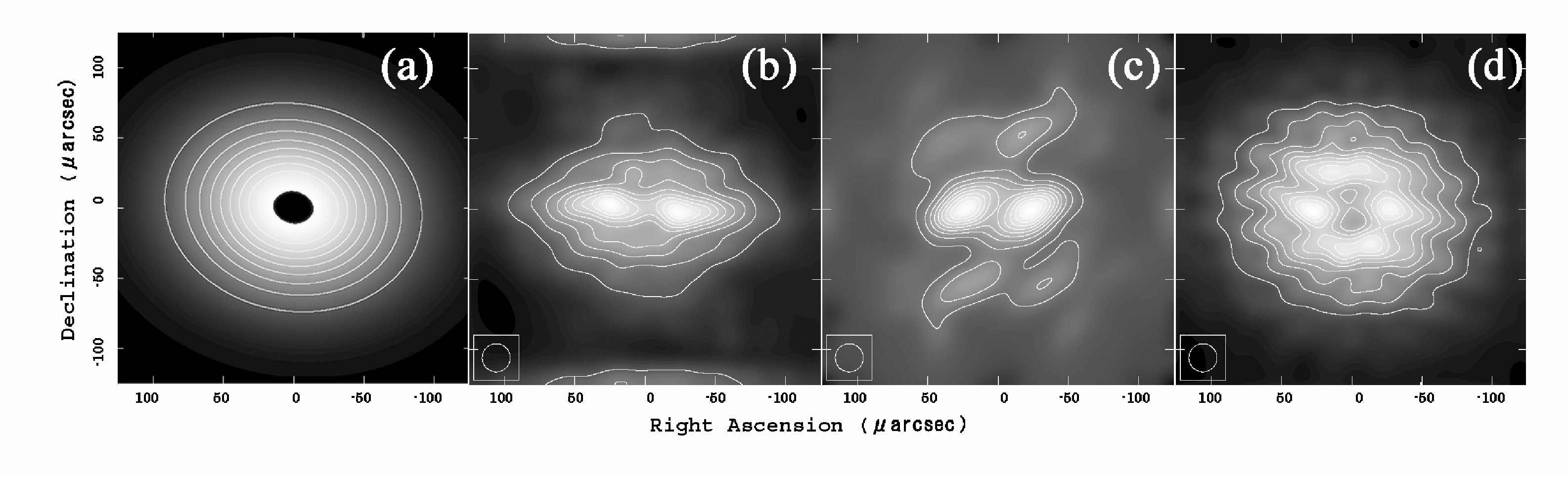}
   \caption{Clean Results from Simulations. The each side is 0.25mas.
(a) Model Image, (b) Image from Array A (VLBA), (c) Image from Array B (FST),
 d)Image from Array C (inverse VLBA).
           }
    \end{figure*}

%
\section{Visibility Analysis}
 Fig. 5 shows visibility curves of three image models, (a) a simple Gaussian brightness without shadow, (b) a Gaussian with the shadow of $30~\mu$ arcsecs. (M$_{BH}=2.6\times10^6M_{\odot})$ and 
(c) a Gaussian with the shadow of $45~\mu$ arcsecs. (M$_{BH}=3.7\times10^6M_{\odot}$).
 For simplicity we used here point symmetric images.
While a Gaussian brightness shows also a Gaussian visibility curve in the amplitude variations,
if the shadow exits the visibility function has null value points at some projected baseline length.
The null value positions changes with the size of shadow. 
From visibility amplitude function, we can distinguish whether the shadow exists or not.
Further, because the null value points move according as the shadow size, we can estimate the shadow size, and the mass of black hole from the null value positions.
 Like old days of VLBI, if the number of baslines is small, visibility analysis will be suitable for investigating the structure of SgrA*

\section{Discussion}

 We here showed only three cases of uv coverage for SgrA*. Much more simulations about uv coverage for SgrA* are described in Miyoshi et. al (2004). In order to get good image of SgrA* black hole, we need a array in Southern hemisphere like VLBA, namely about 10 stations and 8000~km in its extent.

 We note below other technical problems for the observing the black hole in SgrA*.
 
 As for sensitivity, there seems no serious problem. Even the present level of VLBI and sub-mm radio technique gives us the sufficient sensitivity to measure such null value positions in Fig 5. For example, a baseline between two 15-m telescopes with efficiency 0.7, and system temperature 150 K, 1-GHz band width digital recording with quantized efficiency of 0.7, the $1\sigma$ rms noise level reaches 10 mJy with 100 seconds integration (The 3$\sigma$ rms noise level is shown as horizontal red line in figure 5). 

 Aside from the plasma scattering effect, if the accretion disk itself is optically thick at observing wavelength, the black hole shadow will not be seen.  Can we observe the black hole shadow in radio wavelengths?.  Takahashi (2004), however,  shows that the black hole shadow of SgrA* is observable at 230~GHz observations  though the accretion disk is optically think at the frequency (e.g.  Fig. 1) as far as the accretion pattern is not spherical but axis symmetric. 

 The remaining issues are only two. One is the scattering effect by plasma. Here we assumed the effect will become negligible at 230~GHz. While Falcke et al. (2000) estimate the frequency to be 500~GHz. 

 The other is whether we have good observing site in the Southern hemisphere or not. In order to get long baselines we must put VLBI stations away from ALMA site. The separation should be a few hundreds km to a few thousands km.
So we need site survey for sub-mm VLBI stations in the Southern hemisphere. We are planning to start the site survey at Peru in Andes.

 If these two are settled, then we can unveil the black hole environments of SgrA* with a sub-mm VLBI array in the Southern hemisphere.

   \begin{figure*}
   \centering
   \vspace{340pt}
   \includegraphics{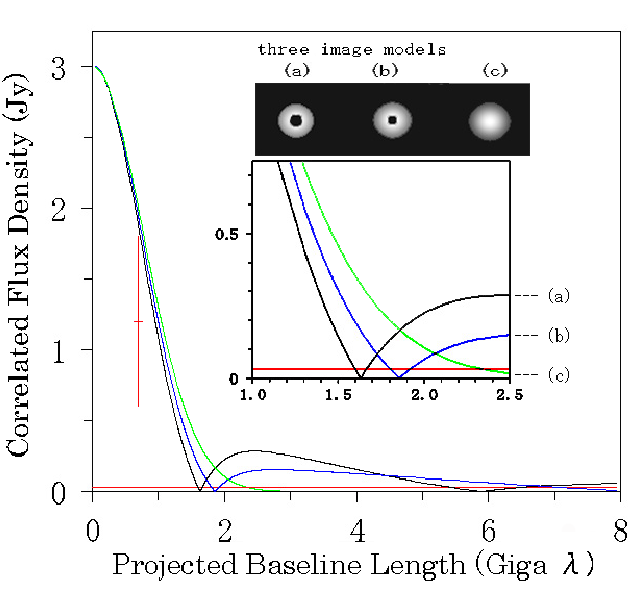}
   \caption{The visibility amplitude functions of three image models with 
projected baseline. (a) the case of $M_{BH}=3.7\times10^{6} M_\odot$, (b) the case of $M_{BH}=2.6\times10^{6}M_\odot$ and (c) the case with no black hole or the scattering effect is still dominant. The functions of (a) and (b) have
 null value points that suggest the real existence of central shadow. The $3\sigma$ noise level of present engineering performance is shown by red horizon line. The point with error bar is measured visibility by Krichbaum et al. (1998)
           }
    \end{figure*}

\begin{acknowledgements}
We would like to thank Mr. Oyama, Dr. Asada for useful discussions.

\end{acknowledgements}

\end{document}